\documentclass[a4paper]{article}

\usepackage{INTERSPEECH2021}
\usepackage{caption}
\usepackage{float}
\title{Speaker Attentive Speech Emotion Recognition}
\name{Clément Le Moine, Nicolas Obin, Axel Roebel}
\address{
  STMS Lab \\
  IRCAM, CNRS, Sorbonne Université,  \\
  Paris, France}
\email{} 

\begin{document}

\maketitle
\begin{abstract}

Speech Emotion Recognition (SER) task has known significant improvements over the last years with the advent of Deep Neural Networks (DNNs). However, even the most successful methods are still rather failing when adaptation to specific speakers and scenarios is needed, inevitably leading to poorer performances when compared to humans. In this paper, we present novel work based on the idea of teaching the emotion recognition network about speaker identity. Our system is a combination of two ACRNN classifiers respectively dedicated to speaker and emotion recognition. The first informs the latter through a Self Speaker Attention (SSA) mechanism that is shown to considerably help to focus on emotional information of the speech signal. Experiments on social attitudes database Att-HACK and IEMOCAP corpus demonstrate the effectiveness of the proposed method and achieve the state-of-the-art performance in terms of unweighted average recall.

\end{abstract}

\section{Introduction}

\subsection{Context}

Speech is one of the most important medium for human communication, not only conveying linguistic information but also rich and subtle para-linguistic information. In particular the affective prosody of speech conveys the emotions which play an important role in the perception of an utterance meaning. Therefore, speech emotion recognition (SER), which aims to recognize the actual emotional state of a speaker from the utterance he produced, has raised a great interest among researchers.

As pointed out by Scherer in \cite{Scherer1986VocalAE}, there is an undeniable contradiction between the apparent ease with which listeners judge emotions from speech and the intricacy of finding discriminative features in speech signal for emotion recognition. This contradiction, in part, lies in the role that individual difference variables play in the production of emotional speech \cite{Bachorowski1999}. In particular, the acoustic characteristics of the speaker voice as well as its speaking style, for a significant part, determine its way of expressing the emotions.

\subsection{Related works}

Formerly adressed using statistical methods and traditional learning techniques such as Hidden Markov Models (HMMs), Gaussian Mixture Models (GMMs) and Support Vector Machines (SVMs), the SER task has known significant improvements over the past years with the advent of deep neural networks (DNNs). Indeed such deep networks have shown excellent abilities to model more complex patterns within speech utterances by extracting high-level features from speech signal for better recognition of the emotional state of the speakers.

Mao et al. \cite{Mao2014} firstly introduced Convolutional Neural Networks (CNNs) for the SER task and obtained remarkable results on various datasets by learning affective-salient features. Recurrent neural networks (RNNs) has also been introduced for SER purpose with a deep Bidirectional Long Short-Term Memory (BLTSM) network proposed by Lee et al. \cite{Lee2015}. Several papers have then presented CNNs in combination with LSTM cells to improve speech emotion recognition, based on log Mel filter-banks (logMel) \cite{Keren2016} or raw signal in an end-to-end manner \cite{Trigeorgis2016}.

Recently, attention mechanisms have raised great interest in the SER research area for their ability to focus on specific parts of an utterance that characterize emotions. Mirasmadi et al. \cite{Mirsamadi2017} approached the problem with a reccurent neural network and a local attention model used to learn weighted time-pooling strategies. Neumann et al. \cite{Neumann2017} used an attentive CNN (ACNN) and showed the importance of the model architecture choice against the features choice. Ramet et al. \cite{Ramet2018} presented a review of attention models on top of BLSTMs and proposed a new attention computed from the outputs of an added BLSTM layer. Chen et al. \cite{Chen2018} proposed a 3-D Attention-based Convolutional Recurrent Neural Networks (ACRNN) for SER with 3-D log-Mel spectrograms (static, deltas and delta-deltas) as input features. They showed 3-D convolution can better capture more effective information for SER compared with 2-D convolution. Recently, Meng et al. outperformed this method by using dilated convolutions in place of a pooling layer and skip connection. 

Attempts to inform emotion classification networks with extra-information involved in the description of emotions were proposed in the past years. Based on previous works \cite{Ververidis2004, Vogt2006, Zhang2018}, Li et al. \cite{Li2019} proposed a multitask learning framework that involves gender classification as an auxiliary task to provide emotion-relevant information leading to significant improvements in SER. Analogously, speaker identity has been used to inform emotion classification networks. The problem was approached by Sidorov et al. \cite{Sidorov2014} with speaker dependent models for emotion recognition. Recently, a method for speaker aware SER was introduced by Assunção et al. \cite{Assuncao2020}, a CNN model VGGVox \cite{Nagrani2017} is trained for speaker identification but is instead used as a front-end for extracting robust features from emotional speech. These first attempts have shown that teaching SER systems with additional signal-based information can greatly improve performances.

\subsection{Contributions of the paper}

In this paper, we assume that individuals may use different means to express emotions, and then that SER should be conditionned on the speaker identity information. Following this hypothesis, we propose a novel method based on previous work \cite{Chen2018}. The major contributions of this paper are summarized as:


\begin{enumerate}
    \item Conditioning emotion classification to speaker identity by using a key-query-value attention we called Self Speaker Attention (SSA) that allows to compute both self and cross-attribute (relation between speaker identity and emotions) attention scores to focus on emotion relevant parts of an utterance.
    \item Proposing a novel regularization by constraining the weights of the last fully connected layer of the network so as to avoid class collapse.
    \item Achieving an absolute increase of UAR by 13.10 \% for IEMOCAP compared with the best existing result.
\end{enumerate}

\section{Proposed Methodology}

In this section, we introduce our SSA-CRNN proposal for SER and a regularization answering class collapse issue. First, the computation of log-Mel spectrograms is described \ref{logmels}, then the basic ACRNN architecture from which our proposal is derived is presented in \ref{ACRNN}. Finally our contributions are detailed : the SSA-CRNN architecture is presented in \ref{SSA-CRNN} and our novel regularization in \ref{reg}. 

\begin{figure*}[h!]
    \centering
    
    \includegraphics[width=\textwidth]{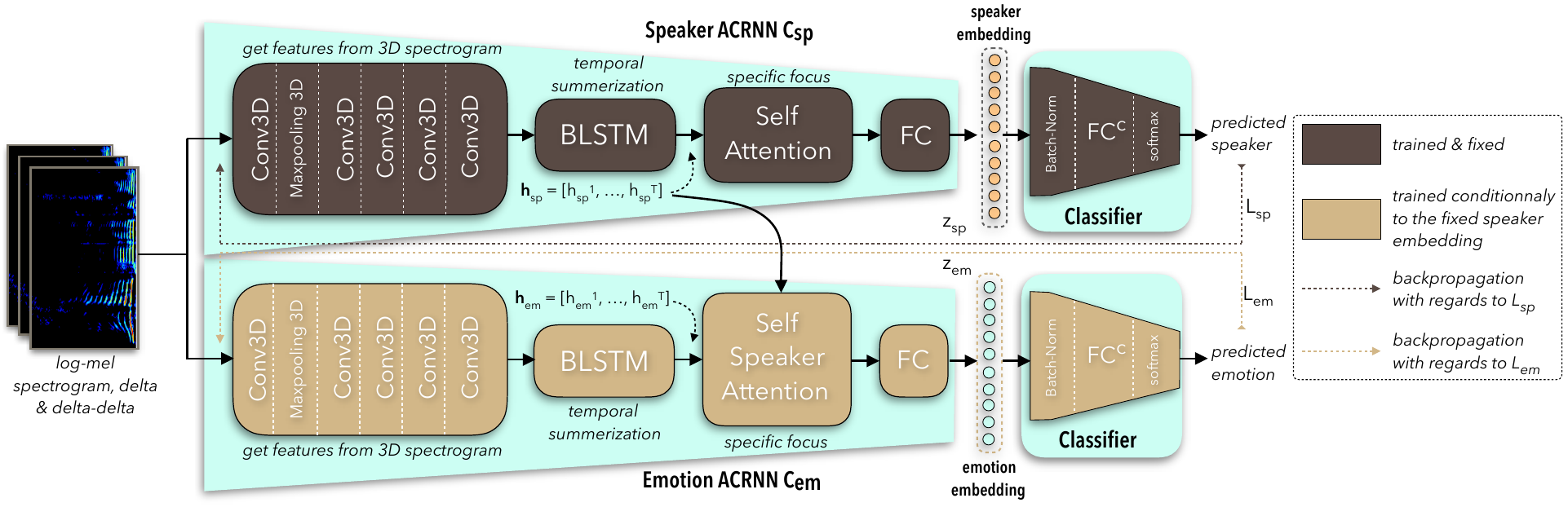}
    \caption{Self Speaker Attentive Convolutional Recurrent Neural Network (SSA-CRNN) for speaker attentive speech emotion recognition composed with two classifiers $C_{sp}$ (up) and $C_{em}$ (bottom) connected through the Self Speaker Attention (SSA) mechanism.}
    \label{SSA-CRNN}
\end{figure*}


\subsection{3-D Log-Mel spectrograms}
\label{logmels}

3-D Log-Mel spectrograms (with delta and delat-deltas), already used as input features for various tasks, were introduced for the SER task by Chen et al. \cite{Chen2018} as input of their ACRNN model and later used in \cite{Meng2019}. In this paper, the Log-Mel spectrograms are computed as presented in \cite{Meng2019}. The 3-D Log-Mel spectrogram consists of a three channel input. The first channel is the static of the Log-Mel spectrogram from 40 filterbanks, the second and third channels are respectively deltas and delta-deltas which can be considered as approximations of the first and second derivatives of the first channel. Once obtained, each 3-D input sample is normalized to have zero mean and unit variance for a given speaker.

\subsection{Architecture of (self) Attentive Convolutional Recurrent Neural Net (ACRNN)}
\label{ACRNN}

Given 3-D log-Mel spectrograms, CRNN is used in \cite{Chen2018} to extract high-level features for speech emotion recognition. The same architecture is used in our proposal to extract high-level features in both cases of SER and speaker recognition (SR).

\subsubsection{Architecture of CRNN}

The CRNN architecture consists of several 3-D convolution layers, one 3-D maxpooling layers, one linear layer and one LSTM layer. Specifically, the first convolutional layer has 128 feature maps, while the remaining convolutional layers have 256 feature maps, and the filter size of each convolutional layer is $5\times3$ , where 5 corresponds to the time axis, and 3 corresponds to the frequency axis. A max-pooling is performed after the first convolutional with pooling size is $2\times2$. The 3D features are reshaped to 2D, keeping time dimension unchanged and passed to a linear layer for dimension reduction before reaching the recurrent layer. As precised in \cite{Chen2018}, a linear layer of 768 output units is shown to be appropriate. These features are then processed through a bi-directional recurrent neural network with long short term memory cells (BLSTM) [19], with 128 cells in each direction,  for temporal summarization, which allows to get 256-dimensional high-level feature representations.

\subsubsection{Self Attention (SA) mechansim}

With a sequence of high-level representations, an attention layer is employed to focus on relevant features and produce discriminative utterance-level representations for classification, since not all frame-level CRNN features contribute equally to the representation of the attributes to recognize, in both cases of SER and speaker recognition. 

Specifically, with the classifier's BLSTM output $\mathbf{h}_{att} = [h_{att}^{1}, ..., h_{att}^{T}] \in \mathbb{R}^{T \times d}$, a temporal vector $\alpha_{att} \in \mathbb{R}^{T}$, representing the contribution per frame to the attribute to recognize, is computed depending on learnt weights vector $W_{att} \in \mathbb{R}^{d}$. Then $\alpha_{att}$ is used to obtain an utterance-level representation by computing the weighted sum of temporal BLSTM internal states $c_{att}$ often called context vector.

\vspace*{-\baselineskip}

\begin{align}
    \alpha_{att} &= \frac{\exp(\mathbf{h}_{att}W_{att})}{\sum_{t=1}^{T}  \exp(W_{att}h_{att}^{t})} \in \mathbb{R^{T}} \\
    c_{att} &= \sum_{t=1}^{T} \alpha_{att}^{t}h_{att}^{t}
\end{align}

The attention layer is followed by a fully connected layer that determines the embedding size. 

\subsection{Architecture of the proposed Self Speaker Attentive Convolutional Neural Net (SSA-CRNN)}
\label{SSA-CRNN}

Our system shown in Figure \ref{SSA-CRNN} is formed by two classifiers $C_{sp}$ and $C_{em}$. The first is an ACRNN and trained for SR. The second has the same architecture apart from its Self Speaker Attention (SSA) layer and is trained (once $C_{sp}$ has been trained) for SER, conditionally to the speakers embedding produced by $C_{sp}$ and through the SSA mechanism described further. This conditioning of SER to speaker identity is expected to help $C_{em}$ to better capture speakers individual strategies for emotion expression leading to better global SER performances.


\subsubsection{Self Speaker Attention (SSA) mechanism}

\begin{figure}[H]
    \centering
    \includegraphics[width=0.48\textwidth]{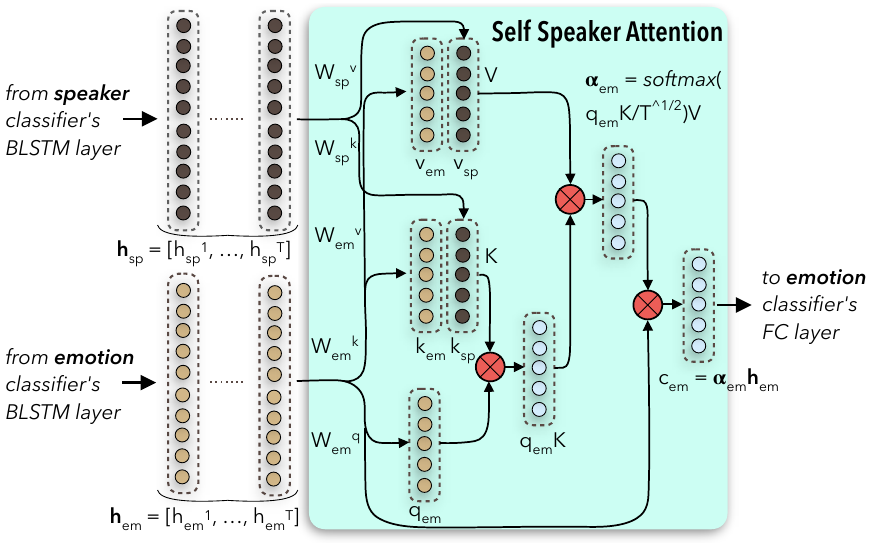}
    \caption{Self Speaker Attention (SSA) Mechanism}
    \label{al}
\end{figure}

Inspired by \cite{Pan2020} that used multi-modal attention technique for SER, we employed the query-key-value representation to compute the attention from two inputs : $\mathbf{h}_{em}$ and $\mathbf{h}_{sp}$ respectively related to "self" and "speaker" aspects. We first compute the query of speech emotion $q_{em}$ through learnable weights $W_{em}^{q} \in \mathbb{R}^{T}$.

\vspace*{-\baselineskip}

\begin{equation}
    q_{em} = \mathbf{h}_{em}W_{em}^{q}
\end{equation}

The keys $K$ and values $V$ are computed using learnable
weights $W_{em}^{k}, W_{sp}^{k}, W_{em}^{v}, W_{sp}^{v} \in \mathbb{R}^{T}$ as follows.

\vspace*{-\baselineskip}

\begin{align}
    K &= concat\{\mathbf{h}_{em}W_{em}^{k}, \mathbf{h}_{sp}W_{sp}^{k}\} \\
    V &= concat\{\mathbf{h}_{em}W_{em}^{v}, \mathbf{h}_{sp}W_{sp}^{v}\}
\end{align}

The cross-attribute and self attention scores are computed by the product of the query $q_{em}$ and transposed keys $K^{T}$. It is then normalized, passed to a softmax activation layer and finally multiplied with values $V$ to obtain $\alpha_{em} \in \mathbb{R}^{T}$ which represents the interaction of speaker identity and speech emotion answering to speech emotion query. This temporal vector $\alpha_{em}$ is then used to compute the weighted sum of temporal BLSTM internal states $c_{em}$.

\vspace*{-\baselineskip}

\begin{align}
    \alpha_{em} &= softmax\left(\frac{q_{em}K^{T}}{\sqrt{T}}\right)V \\
    c_{em} &= \sum_{t=1}^{T} \alpha_{em}^{t}h_{em}^{t}
\end{align}

\subsection{Weights regularization}
\label{reg}

During preliminary experiments with ACRNN on the large database Att-HACK presented in \ref{datasets}, we faced a class collapse issue : the model tended to focus on one or two of the four classes at the expense of the others. That led to have one or several categories very badly recognized. In order to avoid this pitfall, we elaborated a training regularization by constraining the weights $W \in \mathbb{R}^{n_{em} \times 4}$ of the last fully connected layer $FC^{c}$, also called classification layer, just before the \textit{softmax} activation. Denoting $W^{c} \in \mathbb{R}^{n_{em}}$ the c-ieth column of $W$, $N$ the batch size, then for all input batch $\mathbf{x} = [x_{1}, ... , x_{N}]$, $FC^{c}$ is being fed with, the constraint ensuring all classes are equi-outputed  can be expressed as follows.

\begin{equation}
   \lVert W^{c} \rVert_{1} = N\left( 4\lVert\sum\limits_{i=1}^{N}x_{i}\rVert_{1}\right)^{-1}
\end{equation}

\section{Experiments}

\subsection{Datasets}
\label{datasets}

To evaluate the performance of our proposed model, we perform speaker-independent SER experiments on the speech database for speech social attitudes Att-HACK \cite{Le_Moine_2020} and the Interactive Emotional Dyadic Motion Capture database (IEMOCAP) \cite{IEMOCAP}. Att-HACK comprises 20 speakers interpreting  100  utterances in  4  social  attitudes  :  \textit{friendly}, \textit{distant}, \textit{dominant}  and  \textit{seductive}. With 3 to 5 repetitions each per attitude for a total of around 30 hours of speech, the database offers a wide variety of prosodic strategies in the expression of attitudes. IEMOCAP consists of 5 sessions and each session is displayed by a pair of speakers (female and male) in scripted and improvised scenarios. For this experiment, only improvised data and 4 emotions, \textit{happy}, \textit{angry}, \textit{sad} and \textit{neutral} were considered.

\subsection{Experiment Setup}
\label{Experimental Setup}
As for the feature extraction, we split the speech signal
into equal-length segments of 3 seconds for better parallel acceleration and adopt zero-padding for utterances not reaching the 3 seconds.

\begin{itemize}
    \item \textbf{ACRNN} :  Re-implemented strictly following \cite{Chen2018} apart from the embedding size that we increased from $64$ to $128$. Our proposed systems are derived from this baseline SER system $C_{em}$. 
    \item \textbf{ACRNN-r} : Addition of a regularization (described in \ref{reg}) to $C_{em}$ model to avoid class collapse.
    
    \item \textbf{SSA-CRNN-r} : A first ACRNN model $C_{sp}$ (embedding size $64$) is trained for speaker recognition to learn high-level features that define the speaker identity, and from which our SSA mechanism will extract emotion salient information to inform the SER task. Once best accuracy is reached, the model is saved and fixed. Then a second ACRNN model $C_{em}$ (embedding size $128$) is trained conditionally to the BLSTM's outputs of $C_{sp}$ by mean of the SSA mechanism for SER task.
\end{itemize}

For evaluations, 10-fold cross validation technique is performed. Respectively for IEMOCAP and Att-HACK, 9 and 18 speakers are considered for training while the remaining ones are kept for validation. In the case of Att-HACK validation sets are balanced in gender (male-female couples) apart from 2 folds, as the database includes 12 females and only 8 males. As we aspire to perform emotion recognition independently of the speaker, we must consider two approaches for training the SSA-CRNN-r $C_{sp}$ module.

\begin{enumerate}
    \item \textbf{speaker-dependant approach} : a \textit{train/valid} split have been performed linguistically on the whole database, all speakers have been seen by $C_{sp}$. Consequently, the SER task is not performed independently of the speaker and there is no insurance the model will be able to generalize to speakers it has never seen.  
    \item \textbf{speaker-independant approach} : a LOSO (Leave-One-Speaker-Out) protocol is followed so as to ensure that the speaker (or speakers for Att-HACK) on which the model is validated has never been seen by the model. For each fold $k \in [1, 10]$ the speaker $s_{k}^{v}$ (or speakers $[s_{k}^{v_{1}}, s_{k}^{v_{2}}]$) chosen to be in the validation set are left out of the database for the training and validation of $C_{sp}$ .
\end{enumerate} 

Each module is optimized with respect to the cross-entropy objective function. Trainings are done with emotionally balanced mini-batches of 40 samples, using the Adam optimizer with Nestorov momentum. The initial learning rate is set to 0.0001 and the momentum is set to 0.9. The final model parameters are selected by maximizing the Unweighted Average Recall (UAR) on the validation set.

\subsection{Experiment Results}

Table \ref{results} is divided in two parts, the first part shows the results in UAR on IEMOCAP of state-of-the-art methods for SER, including the ACRNN system by \cite{Chen2018} and the gender-informed system by \cite{Zhang2018}. The second part shows the UAR results of our proposed methods on both Att-HACK and IEMOCAP databases. First, we compare our ACRNN re-implementation with the state-of-the-art method described in \cite{Chen2018}. Our re-implementation achieves slightly better performance for IEMOCAP with an absolute improvement of 1.64\% and a rather low standard deviation possibly due to the increased embedding size.

\vspace*{-\baselineskip}

\begin{table}[h!]
    \center
    \begin{tabular}[b]{lcc}
    & \multicolumn{2}{c}{} \\
    \hline
    \textbf{SER system} & \textbf{Att-HACK} & \textbf{IEMOCAP} \\
    \hline \hline
    Chen et al. \cite{Chen2018} & $-$ & $64.74 \pm 5.44$ \\
    Meng et al. \cite{Meng2019} & $-$ & $69.32 \pm 3.76$ \\
    Zhang et al. \cite{Zhang2018} & $-$ & $82.80 \pm -$ \\
    \hline
    ACRNN & $27.49 \pm 3.26$ & $66.38 \pm 2.72$  \\
    ACRNN-r & $66.36 \pm 5.68$ &  $86.35 \pm 3.21$ \\
    SSA-CRNN-r & $88.93 \pm 4.36$ & $96.12 \pm 4.27$ \\
    SSA-CRNN-r (LOSO) & $\mathbf{88.31} \pm \mathbf{5.60}$ & $\mathbf{95.90} \pm \mathbf{4.31}$ \\
    \end{tabular}
    \caption{SER results with 95\% confidence interval for different methods on Att-HACK and IEMOCAP in terms of UAR}
    \label{results}
\end{table}

\vspace*{-\baselineskip}

\begin{figure}[!htbp]

\begin{minipage}[c]{.49\linewidth}
     \begin{center}
             \includegraphics[width=2.5cm]{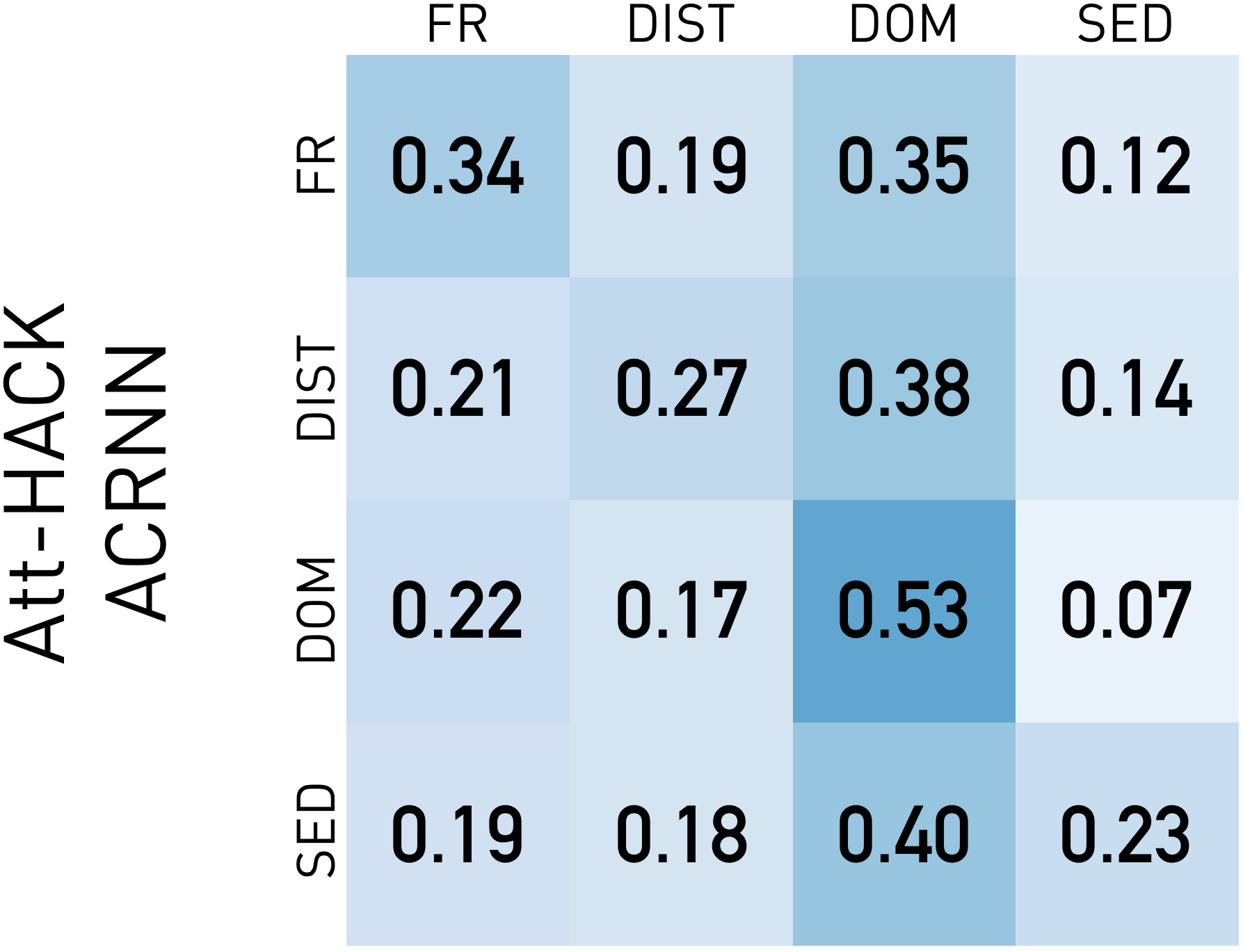}
         \end{center}
   \end{minipage} \hfill
   \begin{minipage}[c]{.49\linewidth}
    \begin{center}
            \includegraphics[width=2.5cm]{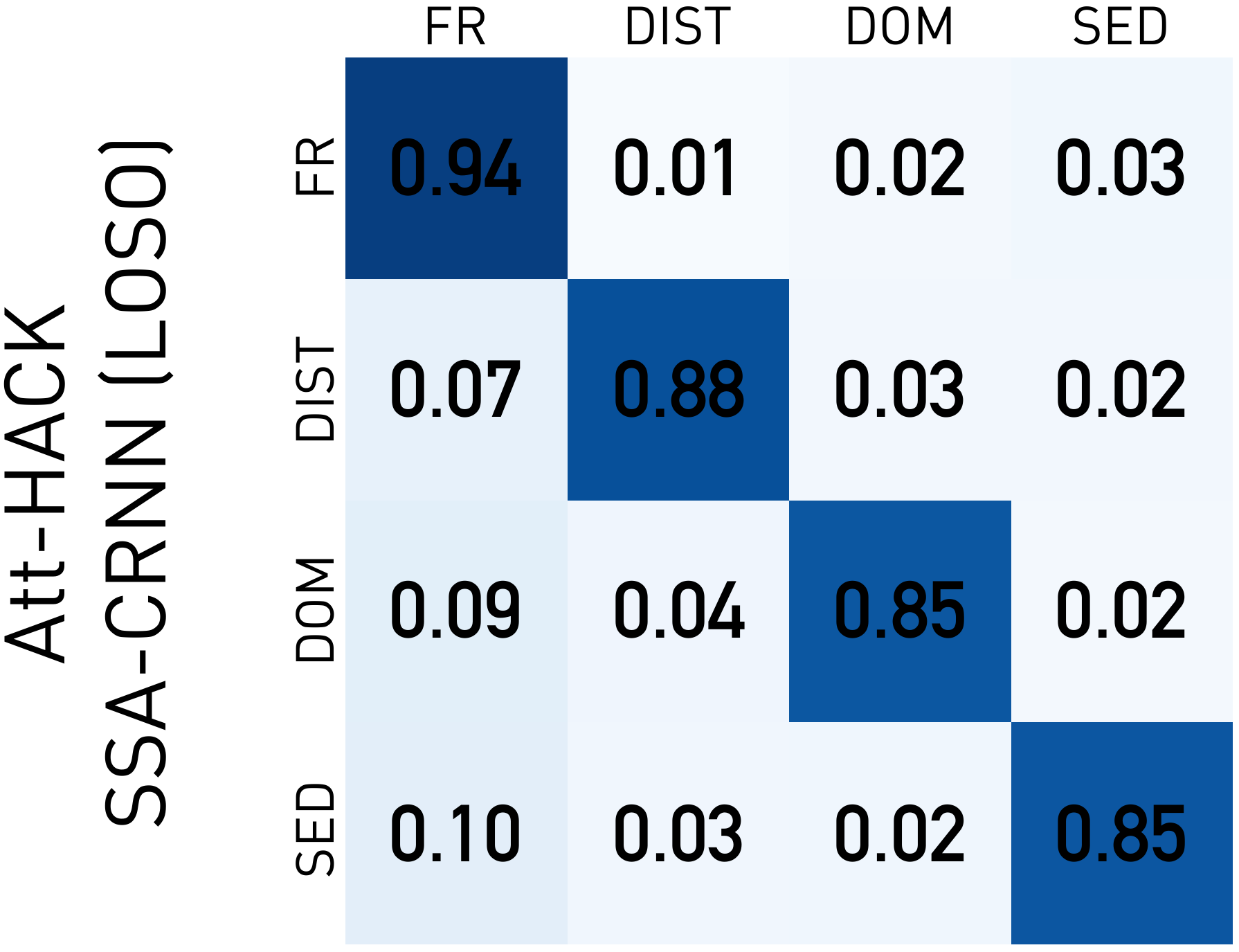}
    \end{center}
 \end{minipage}
\begin{minipage}[c]{.49\linewidth}
     \begin{center}
             \includegraphics[width=2.5cm]{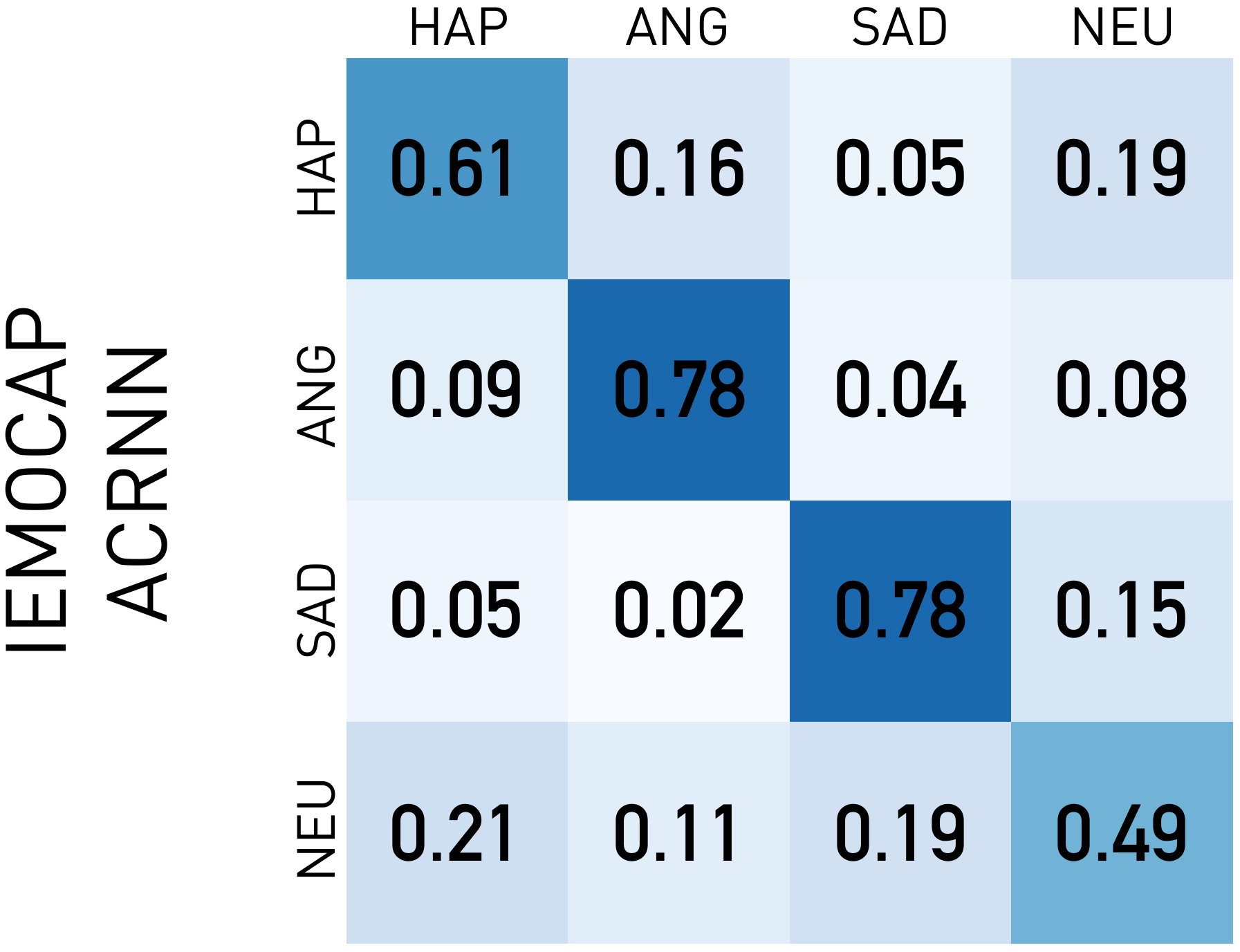}
         \end{center}
   \end{minipage} \hfill
   \begin{minipage}[c]{.49\linewidth}
    \begin{center}
            \includegraphics[width=2.5cm]{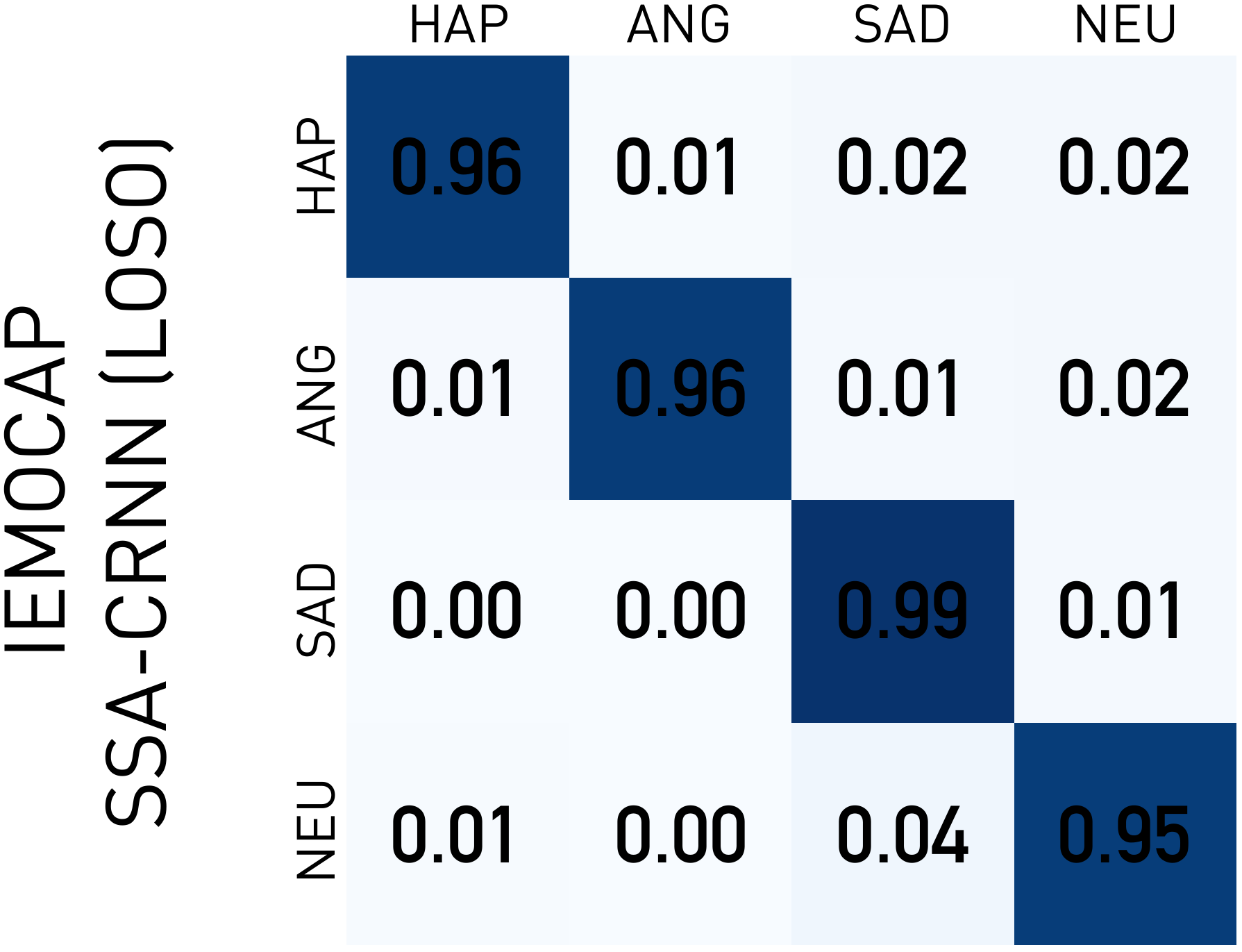}
    \end{center}
 \end{minipage}
 \captionof{figure}{SER normalized confusion matrices where each row presents the class dependant prediction performance.}
 \label{cm}
\end{figure}

Next, we investigate the effectiveness of our proposed regularization. Compared with ACRNN, ACRNN-r obtains an absolute improvement of 38.47\% on Att-HACK and 19.97\% for IEMOCAP. This shows our regularization brings stability during training that allows the model to further recognize more subtle emotions. The confusion matrices depicted in Figure \ref{cm} also show the model is no longer over-focusing on one class, all the other classes are confused, with which was the case for dominant attitude in Att-HACK with the ACRNN. 

To evaluate the addition of our proposed SSA mechanism, it is legitimate to chose the approach with LOSO protocol described in \ref{Experimental Setup}. Compared with ACRNN-r, this approach achieves an absolute improvement of 21.95 \% for Att-HACK and 9.55 \% for IEMOCAP. This shows combining emotion classification and speaker recognition through our SSA mechanism is relevant and lead to better capture speaker means of emotional expression. This approach achieves a global absolute improvement of 60.82 \% for Att-HACK and 29.52 \% for IEMOCAP compared with ACRNN. Compared with the gender-informed SER system by Zhang et al. \cite{Zhang2018}, it achieves 13.10 \% of UAR improvement on IEMOCAP. As expected, the speaker-dependant approach achieves slightly better performance than the LOSO-based one although it does not seem significant. It appears that $C_{sp}$ is able to generalize enough for new speakers so as the SSA can capture the speaker high-level features that are relevant in the representation of speech emotions.  

\vspace{-2.6mm}

\begin{figure}[!htbp]
\begin{center}
        \includegraphics[width=0.47\textwidth]{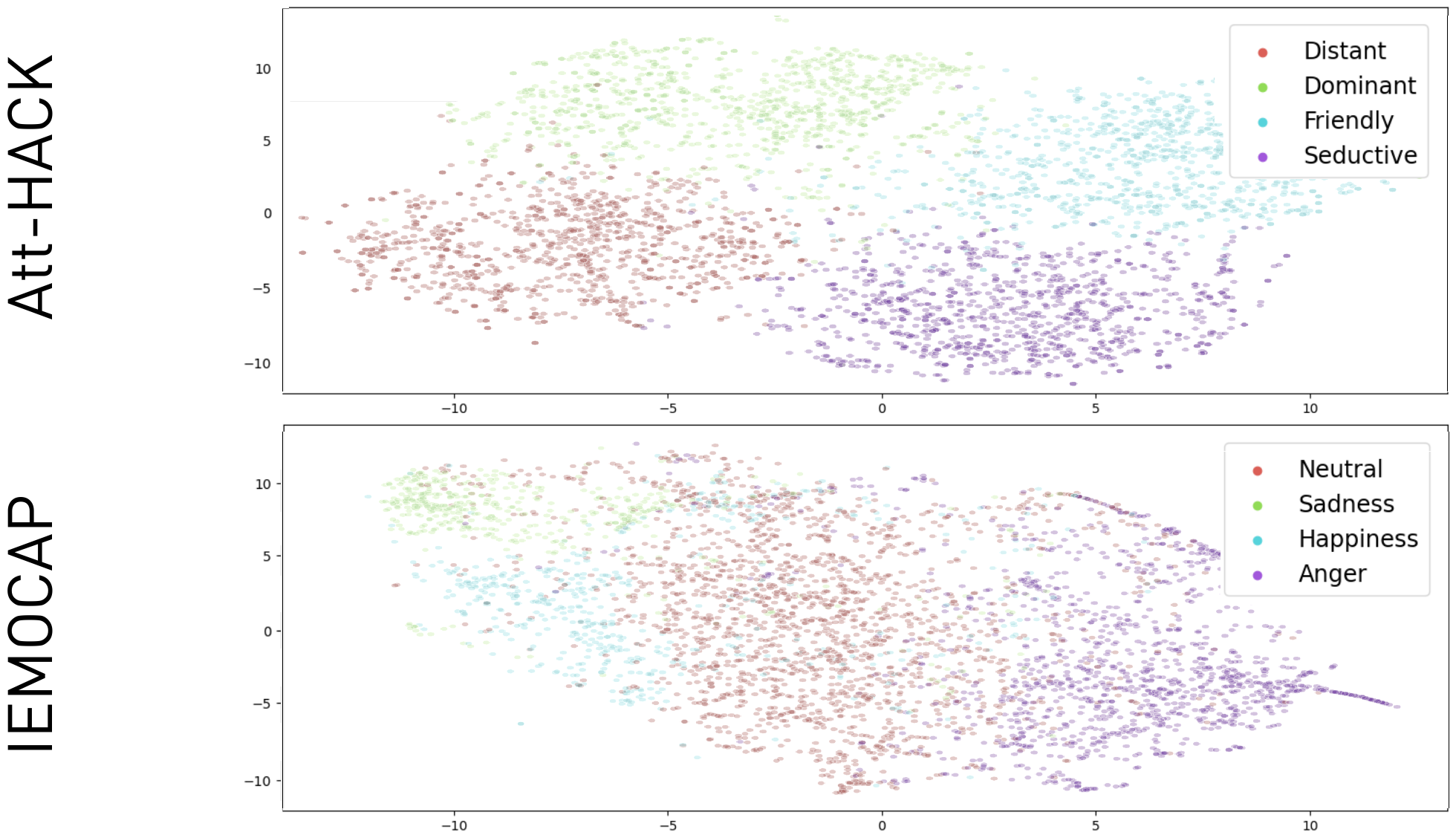}
\end{center}
\caption{T-SNE representations obtained from emotion embeddings, for SSA-CRNN-r (LOSO).}
\label{tsne}
\end{figure}

Finally, the t-SNE analysis in Figure \ref{tsne} tend to show both attitudes and emotions are well separated in their respective embeddings. Although this point would need to be deeper investigated, we can further expect to use these embeddings in a voice conversion context as proposed by Zhou et al. in \cite{Zhou2021}.


\vspace{-2mm}

\vspace*{-\baselineskip}

\section{Conclusions}

In this paper, we propose a Self Speaker Attentive Convolutional Neural Network (SSA-CRNN) model to train the SER model operating on 3D Log-Mel spectrograms and a novel training regularization that allows to avoid class collapse. The evaluation on Att-HACK and IEMOCAP databases demonstrates that the proposed method outperforms the state-of-the-art methods. Notably, the evaluation on IEMOCAP achieves 13.10 \% absolute improvement of UAR compared with the gender-informed SER system by Zhang et al. \cite{Zhang2018}.


\section{Acknowledgements}
\label{sec:akn}
This research is supported by the MoVe project: ‘‘MOdelling of speech attitudes and application to an expressive conversationnal agent'', and funded by the Paris Region Ph2D grant.

\bibliographystyle{IEEEtran}

\bibliography{mybib}


\end{document}